\documentclass[preprint]{aastex}
\usepackage{mkfig}

\begin{document}

\title{\bf Observational Estimates for the Mass-Loss Rates of
  $\alpha$~Centauri and Proxima Centauri Using HST Ly$\alpha$
  Spectra\altaffilmark{1}}

\author{Brian E. Wood,\altaffilmark{2} Jeffrey L. Linsky,\altaffilmark{2}
  Hans-Reinhard M\"{u}ller,\altaffilmark{3} and Gary P. Zank\altaffilmark{3}}

\altaffiltext{1}{Based on observations with the NASA/ESA Hubble Space
  Telescope, obtained at the Space Telescope Science Institute, which is
  operated by the Association of Universities for Research in Astronomy,
  Inc.\ under NASA contract NAS5-26555.}
\altaffiltext{2}{JILA, University of Colorado and NIST, Boulder, CO
  80309-0440; woodb@casa.colorado.edu, jlinsky@jila.colorado.edu.}
\altaffiltext{3}{Bartol Research Institute, University of Delaware, Newark,
  DE 19716; mueller@bartol.udel.edu, zank@bartol.udel.edu.}

\begin{abstract}

     We study H~I Ly$\alpha$ absorption observed by the {\em Hubble Space
Telescope} toward the nearby binary system $\alpha$~Cen (G2~V+K0~V) and
its distant companion star Proxima~Cen (M5.5~Ve).  Absorption from
heliospheric H~I heated by the solar wind/ISM interaction
is observed toward both $\alpha$~Cen and Proxima~Cen.  Absorption from
analogous ``astrospheric'' material surrounding the stars is detected toward
$\alpha$~Cen, but not Proxima~Cen.  The nondetection of astrospheric
absorption toward Proxima~Cen suggests that the stellar wind of Proxima~Cen
must be significantly weaker than that of the $\alpha$~Cen system.  We
use hydrodynamic models of the astrospheres computed assuming different
mass-loss rates to predict astrospheric Ly$\alpha$ absorption for comparison
with the observations.  The model that best matches the $\alpha$~Cen data
has a mass-loss rate of $\dot{M} = 2~\dot{M}_{\odot}$, and the models
suggest an upper limit of $\dot{M} \leq 0.2~\dot{M}_{\odot}$ for Proxima~Cen.
Finally, we note that the heliospheric absorption observed toward
Proxima~Cen in 2000 May is identical to the heliospheric absorption observed
toward $\alpha$~Cen in 1995 May, implying that the structure of the outer
heliosphere does not change significantly during the solar activity cycle.

\end{abstract}


\section{INTRODUCTION}

     Coronal winds analogous to the solar wind have proven to be very
difficult to detect around cool main sequence stars like the Sun.  This
is not surprising given the low density of the solar wind,
corresponding to a mass-loss rate of only about
$\dot{M}_{\odot}=2\times 10^{-14}$ M$_{\odot}$ yr$^{-1}$, and the fact that
the wind is fully ionized.  These properties mean that such winds
cannot be detected spectroscopically like the massive winds of hot stars and
evolved cool stars.

     Ionized winds would be expected to produce radio emission, and
searches for this emission have provided upper limits for the mass-loss
rates of many solar-like stars.  However, these upper limits are typically
3 or more orders of magnitude higher than the solar mass-loss rate
\citep{ab90,sad93,jl96b,ejg00}.  There have been some claims of high
mass-loss rates detected for a few very active stars using observations at
millimeter wavelengths and studies of UV absorption features
\citep{djm89,djm92}.  However, these interpretations of the data remain
highly controversial \citep{jl96a,jl96c}.  There are theoretical arguments
for large mass-loss rates from active stars \citep{gdc76,djm92}, but
\citet{jl96c} argue that massive ionized winds would completely absorb the
coronal radio emission that is commonly observed from these stars.

     Fortunately, a new method for indirectly detecting winds around cool
main sequence stars has recently become available, using spectroscopic
observations of stellar Ly$\alpha$ lines made by the {\em Hubble Space
Telescope} (HST).  Models of the interaction between the solar wind and local
interstellar medium (LISM) predict that charge exchange processes should
create a population of heated neutral hydrogen gas throughout the
heliosphere \citep{vbb95,gpz96,gpz99b,hrm00a}.  This
material produces a detectable absorption signature in the Ly$\alpha$ lines
of many nearby stars \citep{kgg97,vvi99,bew00b}.
However, not only is heliospheric absorption detected in the data, but
analogous ``astrospheric'' absorption from material surrounding the star
is also observed in some cases.

     Astrospheric Ly$\alpha$ absorption has by now been detected for
seven nearby coronal stars, although some of the detections are tentative
\citep{bew96,ard97,kgg97,bew98,bew00a}.  This absorption can only
be present if a stellar wind is present, and if the star is surrounded by
ISM material that is at least partially neutral.  A larger stellar mass loss
rate will result in a larger astrosphere and more
Ly$\alpha$ absorption.  \citet{hrm00b} used this fact to
estimate mass-loss rates for two stars ($\epsilon$~Ind and $\lambda$~And)
based on the amount of observed astrospheric absorption.  In this paper, we
perform a similar analysis to estimate the mass-loss rate of the
$\alpha$~Cen binary system (G2~V+K0~V), which has also been observed to have
detectable astrospheric Ly$\alpha$ absorption \citep{jll96,kgg97}.

     In addition, we present new HST Ly$\alpha$ observations of
$\alpha$~Cen's distant companion star Proxima~Cen (M5.5~Ve), which can be
compared directly with the $\alpha$~Cen data to search for differences in
astrospheric absorption that would indicate differences in stellar wind
properties.  Unlike the two $\alpha$~Cen stars, which will share the same
astrosphere since their orbit has a semimajor axis of only 24~AU
\citep{dp99}, Proxima~Cen will have an astrosphere
all to itself.  Proxima~Cen is about 12,000~AU from $\alpha$~Cen, based on
its $2.2^{\circ}$ separation on the sky and its closer distance of 1.295~pc
compared with 1.347~pc for $\alpha$~Cen \citep{macp97}.  Thus,
while the LISM and heliospheric absorption should be identical toward
$\alpha$~Cen and Proxima~Cen, the astrospheric absorption should be
different.

\section{COMPARING THE LY$\alpha$ LINES OF $\alpha$~CEN AND PROXIMA~CEN}

     Proxima~Cen was observed on 2000 May 8 with the Space Telescope Imaging
Spectrograph (STIS) instrument on HST \citep{bew98a}.
We observed the $1150-1720$~\AA\ spectral range through the
$0.2^{\prime\prime}\times 0.2^{\prime\prime}$ aperture with the moderate
resolution E140M grating.  The total exposure time of the E140M spectrum
was 20,580 s.  The data were reduced in IDL using the STIS team's CALSTIS
software package \citep{dl99}.  The reduction includes assignment of
wavelengths using calibration spectra obtained during the course of the
observations, and a correction for scattered light is performed using the
\verb&ECHELLE_SCAT& routine in the CALSTIS package.  The resulting Ly$\alpha$
spectrum is displayed in Figure~1.  The stellar emission is
contaminated by narrow, weak deuterium (D~I) absorption and very broad,
saturated hydrogen (H~I) absorption.  A geocoronal emission feature is
apparent at 1215.69~\AA, which is removed from the data by fitting a
Gaussian to the feature and then subtracting the Gaussian (see Fig.\ 1).

     Both members of the $\alpha$~Cen binary system were observed in 1995
May with the Goddard High Resolution Spectrograph (GHRS) instrument that
preceded STIS on board HST.  The observations included high resolution
spectra of the Ly$\alpha$ line, which
were analyzed by \citet{jll96}.  The $\alpha$~Cen~B spectrum is
shown in Figure~2, where we have normalized the fluxes using the assumed
stellar emission profile.  Since a precise LISM H~I column density could
not be derived for the $\alpha$~Cen line of sight, \citet{jll96}
actually presented two possible stellar profiles.  We use the one that
results in a derived deuterium-to-hydrogen ratio of
${\rm D/H}=1.5\times 10^{-5}$, the value that \citet{jll98} finds to be
consistent with very nearby ISM material along many lines of sight.  The
LISM H~I column density derived from the $\alpha$~Cen~B data using this
profile is $\log {\rm N_{H~I}}=17.60$.

     We use this column density to estimate the stellar emission line
profile for Proxima~Cen (see Fig.\ 1) using the following procedure.  By
assuming this column density, we can compute a
wavelength dependent opacity profile for the H~I absorption, $\tau_{\lambda}$.
The stellar profile outside the saturated core of the H~I absorption is then
derived simply by extrapolating upwards from the data, multiplying
the spectrum by $\exp (\tau_{\lambda})$.  We then interpolate the profile
over the saturated H~I absorption core.

     The green dashed line in Figure~2 shows the LISM absorption toward
$\alpha$~Cen, based on a fit in which the H~I absorption is forced to
have a central velocity and Doppler broadening parameter consistent with D~I
and other LISM absorption lines \citep{jll96}.  Excess H~I
absorption is apparent on both sides of the LISM absorption.  \citet{kgg97}
used hydrodynamic models of the heliosphere to show that heliospheric
H~I could account for the excess absorption on the red side of the line, but
not the blue-side excess.  The redshift of the heliospheric absorption
relative to the LISM is due to the deceleration of interstellar material as
it crosses the bow shock.  The blue-side excess is presumably due to
analogous ``astrospheric'' material surrounding the star, which is seen as
blueshifted rather than redshifted because we are observing the material
from outside the astrosphere rather than inside \citep{kgg97}.

     The observed Proxima~Cen fluxes are normalized to the stellar profile
shown in Figure~1, and the result is plotted in Figure~2 for comparison
with the $\alpha$~Cen~B data.  We shifted the Proxima~Cen spectrum by
2 km~s$^{-1}$ to force its D~I absorption feature to line up with that of
the $\alpha$~Cen spectrum.  The Proxima~Cen data have a lower spectral
resolution than the $\alpha$~Cen~B data, explaining why the D~I line is
broader and not as deep.  The amount of observed D~I absorption toward the
two stars is essentially identical, as one would expect.

     The Ly$\alpha$ absorption profiles of Proxima~Cen and $\alpha$~Cen
agree very well on the red side of the line, implying that the heliospheric
absorption responsible for the aforementioned red-side excess is identical
toward both stars.  We would not expect to see any spatial variations in
heliospheric absorption between two stars so nearby in the sky, but the data
were taken five years apart and the heliosphere could have conceivably changed
in the interim.  The 1995 $\alpha$~Cen data were taken close to the minimum
of the Sun's activity cycle, while the 2000 Proxima~Cen data were taken
close to solar maximum.  Apparently, the structure of the outer heliosphere
does not vary significantly during the solar activity cycle.  This result is
consistent with the theoretical predictions of \citet{gpz99a}, whose
time-dependent heliospheric models predict little variability for global H~I
properties in the outer heliosphere despite the fact that the solar wind ram
pressure near Earth varies by about a factor of two during the solar
activity cycle \citep{ajl90,jdr97}.

     Unlike the red side of the Ly$\alpha$ line, the Ly$\alpha$ absorption
profiles of $\alpha$~Cen and Proxima~Cen do {\em not} agree well on the
blue side of the line.  The blue
side of the Proxima~Cen Ly$\alpha$ absorption agrees well with the estimated
ISM absorption, meaning there is no detectable astrospheric absorption
toward Proxima~Cen, in contrast to $\alpha$~Cen.  Note that the blue-side
excess absorption seen toward $\alpha$~Cen and other stars has been
interpreted as being due to astrospheres primarily because the properties of
the absorption are consistent with theoretical expectations for astrospheric
absorption.  The difference in absorption between Proxima~Cen and
$\alpha$~Cen seen in Figure~2 represents the strongest {\em purely empirical}
evidence that the blue-side excess absorption observed toward $\alpha$~Cen
is indeed due to circumstellar material surrounding $\alpha$~Cen, which does
not extend as far away as Proxima~Cen.  {\em This provides particularly
strong evidence for the astrospheric interpretation of the excess
absorption.}  Differences in stellar wind properties must be responsible for
the difference in astrospheric absorption.  In particular, the significantly
lower astrospheric H~I column density of Proxima~Cen suggests a much smaller
astrosphere, which in turn suggests that Proxima~Cen's wind is much weaker
than that of $\alpha$~Cen.

\section{ESTIMATING MASS-LOSS RATES}

     In an attempt to estimate mass-loss rates for $\alpha$~Cen and
Proxima~Cen, we compute a series of hydrodynamic models of the astrospheres
assuming different loss rates (see Fig.\ 3).  These models are extrapolated
from a heliospheric model that correctly predicts the amount of heliospheric
absorption for various lines of sight \citep*[see][]{bew00b}.  We assume
the same input parameters for the astrospheric models,
with the following exceptions.  Instead of an ISM temperature of 8000~K
appropriate for the Local Interstellar Cloud (LIC) around the Sun, we assume
a temperature of 5650~K for the ISM surrounding Alpha/Proxima~Cen, since
these stars are not in the LIC, but are in the cooler ``G cloud''
\citep*{jll96,bew00a}.  Based on the G cloud flow vector \citep{rl92}
and the known stellar space motion of the
Alpha/Proxima~Cen system, we find that the stars see a slightly slower ISM
wind speed than the Sun:  25 km~s$^{-1}$ compared with 26 km~s$^{-1}$ for
the Sun.  This slight  difference in ISM flow velocity is also taken into
account.

     The ``baseline'' heliospheric model we are extrapolating from assumes
a proton density for the wind of $n(H^{+})=5.0$ cm$^{-3}$ at 1~AU from the
Sun.  In our astrospheric models, we experiment with different stellar
mass-loss rates by changing $n(H^{+})$.  Figure~3 shows the H~I density
distribution for four models with mass loss rates of $0.2~\dot{M}_{\odot}$,
$0.5~\dot{M}_{\odot}$, $1.0~\dot{M}_{\odot}$, and $2.0~\dot{M}_{\odot}$,
illustrating how the size of the astrosphere increases with increasing mass
loss.  The H~I density enhancement shown in red is the ``hydrogen wall''
between the bow shock and ``astropause'' (analogous to the ``heliopause''),
which will be responsible for most of the astrospheric absorption for the
line of sight toward the Sun, $79^{\circ}$ from the upwind direction.

     The models provide tracings of H~I temperature, density, and projected
flow velocity along the line of sight toward the Sun from which we can
compute the astrospheric Ly$\alpha$ absorption predicted by each model.  The
predicted absorption of the four models is shown in Figure~2.  The
$2.0~\dot{M}_{\odot}$ model reproduces the $\alpha$~Cen data well.  This is
a very sensible result, because the two $\alpha$~Cen stars individually
have coronae with solar-like temperatures and X-ray luminosities \citep{mh99}
but collectively have about twice the surface area of the Sun,
assuming radii of 1.20~R$_{\odot}$ and 0.90~R$_{\odot}$ for $\alpha$~Cen~A
and $\alpha$~Cen~B, respectively \citep*{dp99}.
The $0.2~\dot{M}_{\odot}$ model is the only model that does not
produce too much absorption to be consistent with the Proxima~Cen data, and
therefore represents an upper limit for Proxima~Cen's mass-loss rate
($\dot{M}\leq 0.2~\dot{M}_{\odot}$).  Despite this low value Proxima~Cen's
mass loss per unit surface area could still be as much as 8 times larger
than the Sun, since Proxima~Cen's radius of 0.16~R$_{\odot}$ \citep{pmp93}
implies a surface area about 40 times smaller than that of the Sun.

     We should point out that our mass-loss estimates may contain sizable
systematic errors since our technique of applying solar wind models with
rescaled densities to stellar winds relies on significant assumptions about
the applicability of these models to other stars.  For example, we assume
that the stellar winds have the same velocity as the solar wind at 1~AU
($v=400$ km~s$^{-1}$).  To first order, the size of an astrosphere and the
column density of astrospheric absorption should scale with the square root
of the wind ram pressure, $P_{wind}$ \citep{bew98}.  Since
$P_{wind}\propto \dot{M} v$, our mass loss estimates will vary inversely
with the assumed wind speed.  For the $\alpha$~Cen stars, a wind velocity
equivalent to that of the Sun is clearly the best assumption, since these
stars and their coronae (where the winds are accelerated) are very
solar-like.  However, Proxima~Cen's corona has a significantly higher
temperature and X-ray surface flux than the Sun \citep{mh99},
so its wind velocity is more likely to be different.

     The low mass-loss rate that we find for Proxima~Cen is not surprising
in the sense that Proxima~Cen is much smaller and dimmer than the
$\alpha$~Cen stars and the Sun.  However, like many M dwarf stars it has a
surprisingly active corona that produces large flares, and it has a
quiescent X-ray luminosity about equal to that of the Sun and $\alpha$~Cen
\citep{mh99}.  Since the winds of cool main sequence stars are
accelerated in the corona, one might therefore expect that Proxima~Cen's
wind might be as strong or stronger than that of $\alpha$~Cen and the Sun.
Indeed, it has been proposed that the large flares on active M dwarfs like
Proxima~Cen should induce very large mass loss rates \citep{gdc76,djm92}.
Our observations suggest that this is not the case.
However, general conclusions
about the mass-loss rates of active M dwarfs should await observations of
additional active M stars, especially since Proxima~Cen's activity level is
quite modest compared to many M dwarfs.


\acknowledgments

    Support for this work was provided by NASA grants NAG5-9041 and
S-56500-D to the University of Colorado.

\clearpage

\begin{figure}
\plotfiddle{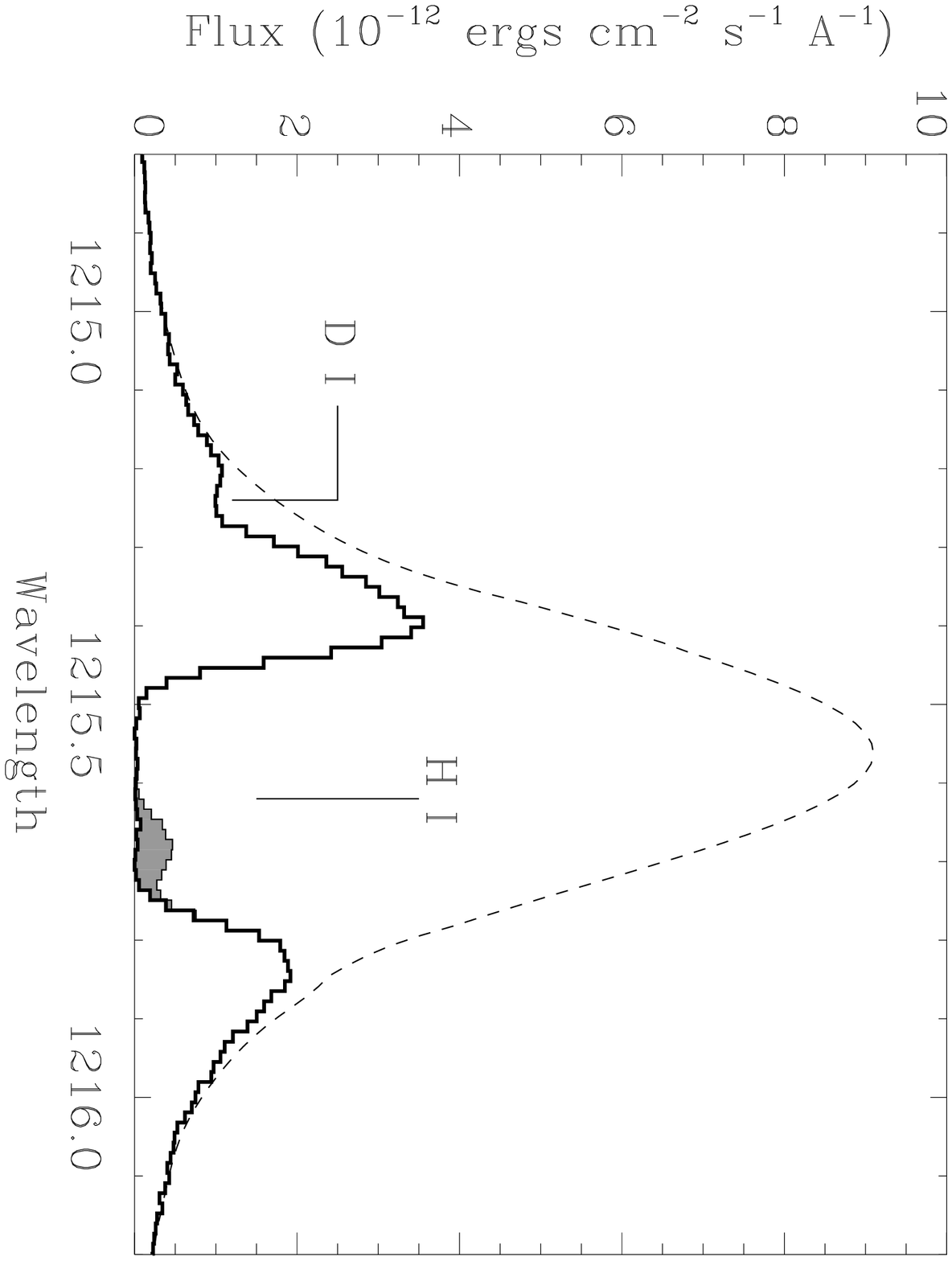}{3.5in}{90}{75}{75}{280}{0}
\caption{The HST/STIS spectrum of the Ly$\alpha$ line of Proxima~Cen
  before and after removal of the geocoronal absorption (shaded region).
  The centroids of the H~I and D~I absorption are indicated.  The dashed line
  is the intrinsic stellar emission line assumed in the analysis.}
\end{figure}

\clearpage

\begin{figure}
\plotfiddle{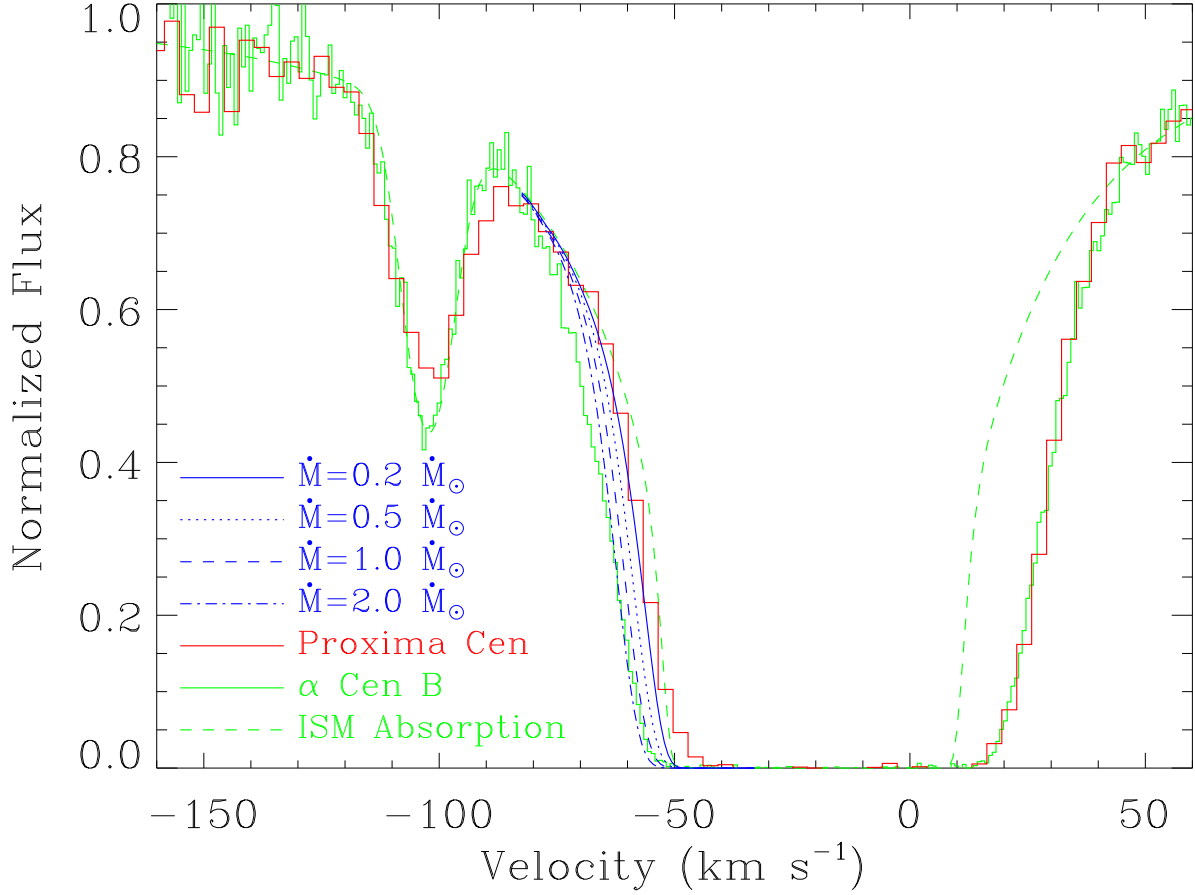}{3.5in}{90}{75}{75}{280}{0}
\caption{Comparison between the Ly$\alpha$ spectra of $\alpha$~Cen~B
  (green histogram) and Proxima~Cen (red histogram).  The inferred ISM
  absorption is shown as a green dashed line.  The Alpha/Proxima Cen data
  agree well on the red side of the H~I absorption, but on the blue side the
  Proxima Cen data do not show the excess Ly$\alpha$ absorption seen toward
  $\alpha$~Cen (i.e., the astrospheric absorption).  The blue lines show the
  blue-side excess Ly$\alpha$ absorption predicted by four models of the
  Alpha/Proxima Cen astrospheres, assuming four different mass loss rates.
  The $2.0~\dot{M}_{\odot}$ model fits the $\alpha$~Cen spectrum reasonably
  well, and the $0.2~\dot{M}_{\odot}$ model represents an upper limit for
  the mass loss rate of Proxima~Cen.}
\end{figure}

\clearpage

\begin{figure}
\plotfiddle{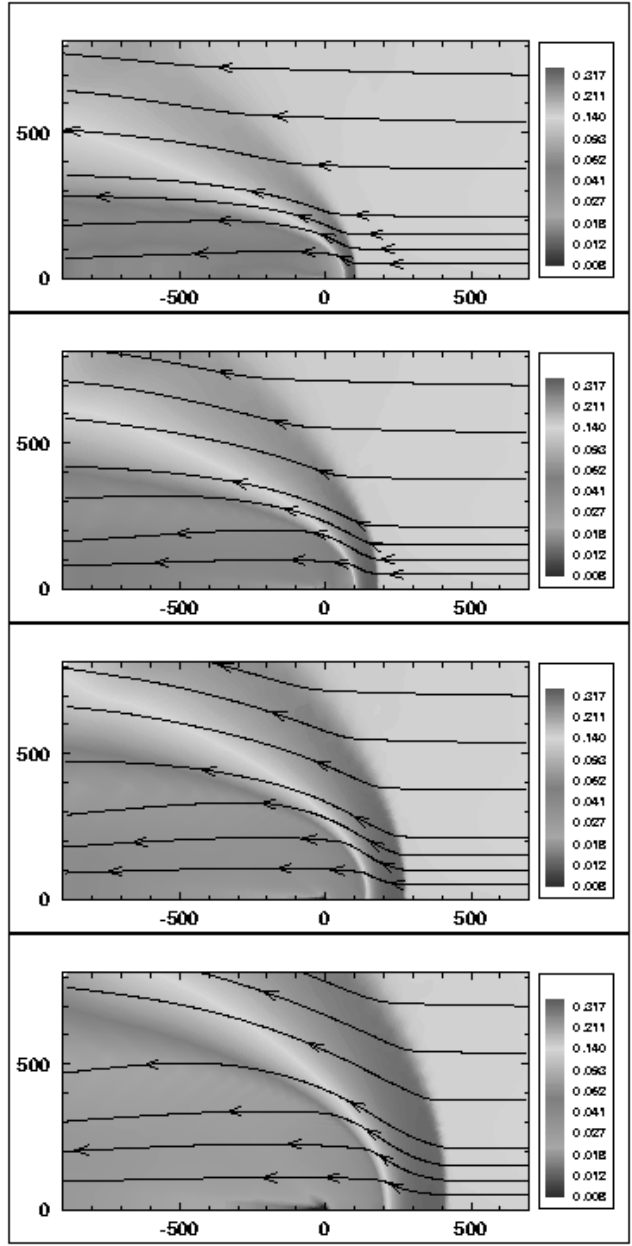}{6.0in}{0}{65}{65}{-150}{-20}
\caption{Distribution of H~I density predicted by hydrodynamic models
  of the Alpha/Proxima~Cen astrospheres, assuming stellar mass loss
  rates of (from top to bottom) $0.2~\dot{M}_{\odot}$, $0.5~\dot{M}_{\odot}$,
  $1.0~\dot{M}_{\odot}$, and $2.0~\dot{M}_{\odot}$.  The distance scale is
  in AU.  Streamlines show the H~I flow pattern.}
\end{figure}

\end{document}